# Anomalous emissions of $^{103m}$Rh biphoton transitions

Yao Cheng[1] and Bing Xia[2]

[1] Department of Engineering Physics, Tsinghua University, 100084, Beijing, China

[2] Institute of Nuclear and New Energy Technology, Tsinghua University, 100084, Beijing, China

yao@Tsinghua.edu.cn

Abstract

In this report, the anomalous emissions, centered on the one half transition energy ~ (39.76/2) keV, are observed from the long-lived Mössbauer state of $^{103m}$Rh excited by bremsstrahlung exposure. Strong coupling with identical nuclei in Rh crystals opens cascade channels for biphoton transitions.

## 1. Introduction

Recently, we reported on the Mössbauer effect of extraordinary long-lived $^{103m}$Rh states excited by bremsstrahlung exposure at 300 K and at 77 K [1-4]. This work reports on the anomalous emissions (AE) from cascade channels [2] of simultaneously emitted biphotons [5, 6]. In the strong-coupling regime, the biphoton emission and re-absorption process within a crystal consisting of identical resonators is entirely recoilless, the process being able to transfer energy and momentum via a three-body entanglement of a biphoton and phonon [2]. Accordingly, a massive neutral quasiparticle of the nuclear spin-density wave (NSDW) is suggested as a means to interpret the quantum phase transition [1, 2].

Dynamic Bose-Einstein condensation (BEC) of the quasiparticle magnon has been demonstrated under pumping at room temperature [7-10]. The pumping efficiency is enhanced by matching two short-wavelength magnons of opposite wavevectors with a long-wavelength microwave, where the parametrically-generated magnons have frequencies centered at one half the microwave pumping frequency. Moreover, to detect the radiation from condensate using an antenna with a size much larger than the magnon wavelength is also enhanced by two-magnon confluence [10]. The biphoton emission and re-absorption reported here yields similar accounts [2], wherein two γs with opposite wavevectors maintain their strong coupling with nuclei on lattice sites. In contrast to a straightforward BEC of bosonic magnons [7-10], the biphoton NSDW when excited beyond a critical density most likely undergoes a pseudo-BEC with Cooper pairing [11], as discussed in previous works [1, 2].

Multipolar AE from an internal Mössbauer emitter have long been predicted by Hannon and Trammell using the reciprocity theorem [12]. With E-field nodes but B-field antinodes at lattice sites, AE decouple from electrons but are strongly coupled to multipolar emitters. A potential γ laser was discussed [12]. Instead of a single-photon AE, we suggest biphoton AE. These biphoton AE show up within three Mössbauer nuclides, $^{103}$Rh, $^{93}$Nb and $^{45}$Sc, which are single isotopes with multipolar transitions. Biphoton γ lasing may emerge from a coherent NSDW condensate.

The AE reported in this work have a broad-band structure, of which the relative intensity for one incoming γ count is fixed in three excitation regimes [1-4]. We shall show that AE are emitted from $^{103m}$Rh. We found other kinds of AE, which shall be detailed elsewhere; for example, three narrow-banded AE peaks located at 17.4, 19.9 and 22.4 keV show up along the long edge. The anisotropy, which depends on the macroscopic sample geometry, reveals the NSDW texture, as has been briefly discussed in [1, 2]. Cooling samples to 77 K significantly enhanced the central AE at 19.9 keV along the long edge that supports two different biphoton modes [2].

## 2. Measurements

The experimental procedure, the polycrystalline sample (25 × 25 × 1 mm$^3$) and the detection system are the same as those reported previously [1], wherein the three data points in regime III, identified as 1, 2, 3, are again selected with filters inserted between the sample and the detector, as shown in figure 1 in Ref. [3]. These data points were obtained in the absence of magnetic fields. Data point 1 were measurements performed without filters, data point 2 with a Cu filter, and data point 3 with a Ta filter. The copper foil has a thickness of 35 μm, while the tantalum foil has a thickness of 25 μm. The energy spectra are taken with a channel width of 25.7 eV.

## 3. Data analysis

Coincidence counts associated with two photons entering a detector within a time span much shorter than its temporal resolution are counted as a single photon [13]. Such counts cannot be rejected by the system's electronics. Peak pile-ups appear at the sum of the energies of two randomly incident photons. Pile-ups from $^{103m}$Rh thus have a double decay speed, *i.e.* 2/4857 s$^{-1}$. However, coincidence counts show a single decay speed when two incoming photons are temporally entangled [1, 2]. In this work, we carefully identify by means of their decay speeds these two different events as peak pile-ups and coincidence counts. To remove the extra counts associated with peak pile-ups, we applied off-line analysis to eliminate "double decay speed" counts. The time-resolved spectral profiles taken over a 3 hour duration are separated into two unequal periods, denoted by $S_1(E)$, and $S_2(E)$. Here $E$ is the energy of the spectral profiles. Data from the first period was gathered over an initial 28 min duration ($T_1$), which corresponds to the 1/2 half-life of $^{103m}$Rh, while the second period corresponds to data gathered over the remaining measurement time $T_1$-$T$. The full spectral profile $S(E)$ is calculated by the formula

$$S(E) = aS_2(E) - S_1(E), \qquad (1)$$

with the coefficient $a$ obtained from

$$\int_0^{T_1} e^{-2t/\tau} dt - a \int_{T_1}^{T} e^{-2t/\tau} dt = 0, \qquad (2)$$

where $\tau$ = 4857 s is the known decay time constant of $^{103m}$Rh, and $T$ = 3.064 hours is the total data-taking duration. That $T$ fractionally exceeds a 3-hour duration is due to the idle time arising in transferring data from the detecting system to the computer. According to Eq. (1), we obtain $a$ = 1.049. The off-line analysis involves saving counts with decay time constant of 4857 s, and removing any pile-up counts decaying with time constant of 4857/2 s.

From the energy spectra of $^{103m}$Rh, shown in figure 1(a), the Kα and Kβ peaks are easily identified. The peaks at 28.8 and 29.9 keV are the escape peaks of the 39.76-keV γ photons. The three curves presented in figure 1(a), are from top to bottom respectively for spectra taken without filters (curve 1 in red color for the on-line version), with a Cu filter (curve 2 in blue), and with a Ta filter (curve 3 in green). The full spectral profiles $S(E)$, including curve 1 without filters and curve 2 with a Cu filter as shown in figure 1(a), are calculated by (1) and normalized to one incoming γ. Due to the strong suppression of the K counts by the Ta filter by over a factor of 20, curve 3 is presented only by normalization without off-line analysis using (1).

The Compton edge for the γ peak of 39.76 keV is approximately 5.4 keV, which is far below the range of interest. For the spectra from $^{109m}$Ag ($^{109}$Cd source) presented in figure 1(b), the escape peaks of the 88.03-keV γ and the characteristic K lines from the lead shielding are identified within the energy range from 72 to 85 keV. The Compton edge corresponding to the γ peak of 88.03 keV is approximately 22.5 keV, which is not shown in figure 1(b). A "bump" in the spectrum shows up from 65 to 88 keV that is attributed to the Compton back scattering of the 88.03 keV γ-ray of $^{109m}$Ag ($^{109}$Cd) from the passive material installed behind the detector by the manufacturer [14]. The decreasing intensity extending from 65 down to 52 keV is due to double Compton scattering events. The "valley" in the intensity of order 10$^{-5}$ between the Compton edge and the double scattering shelf in figure 1(b) is therefore attributed to the intrinsic detector properties [13, 14]. A similar "valley" is expected with a comparable intensity of 10$^{-5}$ from the double scattering shelf at 30 keV extending down to the Compton edge at 5.4 keV for the γ peak of $^{103m}$Rh at 39.76 keV. However, a plateau-like continuum shows up in this region with intensity larger than the value expected for this "valley" by at least one order of magnitude. In addition, a "bump" similar to that appearing at ~65 keV in figure 1(b) is also expected in the spectrum of $^{103m}$Rh at about 35 keV as a consequence of Compton back scattering. However, due to the presence of the plateau-like anomalous continuum, the structure of the expected "bump" appears more as a "shoulder" with an emission intensity of $10^{-3}$ to $10^{-4}$ from 35 to 40 keV in figure 1(a), which is comparable to that from 65 to 88 keV in figure 1(b).

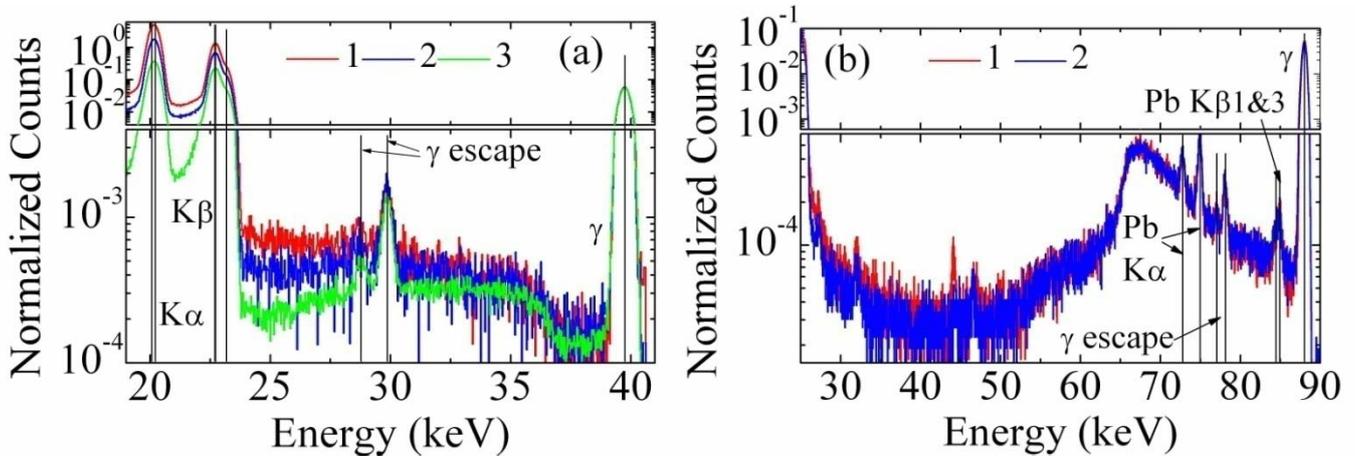

**Fig.1.** (Colour on line) Normalized spectra for (a) the $^{103m}$Rh sample and (b) the $^{109m}$Ag from nuclear transmutation of the $^{109}$Cd source by electron capture. The spectral profiles are normalized to the total count of the incoming γ of energy 39.76 keV for $^{103m}$Rh and 88.03 keV for the $^{109m}$Ag ($^{109}$Cd source). For the spectral profiles of $^{103m}$Rh presented in (a), the energy range of interest is from the Kα peak, slightly less than 20 keV, to the γ peak above 40 keV. Curve 1 in red (on the top) is obtained without filters. Curve 2 in blue (at the bottom) is obtained with a copper filter. Curve 3 in green is obtained with a tantalum filter. Curves 1 and 2 are calculated according to (1), whereas curve 3 is the experimental data. For the spectral profiles of $^{109m}$Ag ($^{109}$Cd source) presented in (b), the energy range is from 25 keV to 90 keV slightly above the γ peak. Due to the low count rates, the Ag spectra are presented with the experimental data without off-line analysis of (1). Curve 1 in red is the spectrum obtained without filters, whereas curve 2 in blue is with a copper filter. The two curves almost coincide. The four peaks from 70 to 80 keV are the Kα1 and Kα2 from the Pb shielding and γ escapes from $^{109m}$Ag. The two peaks around 85 keV are the Kβ1 and Kβ3 of the Pb shielding, while Kβ2 at 87.3 keV is buried in the γ peak.

Aside from the two escape peaks at 28.8 and 29.9 keV in the spectra of $^{103m}$Rh, seen in figure 1(a), there exists a plateau-like broad-band continuum from 24 to 35 keV between the Kβ and γ peaks, as discussed near the end of the preceding paragraph. In this region, the intensity almost reaches a level of $10^{-3}$ relative to the single incoming γ count and the energy is much higher than the Compton edge at 5.4 keV for the γ of 39.76 keV emitted from $^{103m}$Rh. The experimentally observed intensity is thus much higher than the expected value of order $10^{-5}$ estimated according to the intensity level of the "valley" in figure 1(b) because of nearly equal Compton cross-sections for both 40 and 88 keV [15]. In figure 1(b), the intensity of the spectrum for $^{109m}$Ag ($^{109}$Cd) remains unaffected by insertion of the copper filter. This is as expected because the detected counts are an intrinsic property of the detector with respect to the 88.03-keV γ ray. However, the level of the broad-band emissions of $^{103m}$Rh is attenuated by inserting Cu and Ta filters, as shown in figure 1(a). It indicates that the plateau-like broad-band feature is not attributed to any detector effect associated with the incoming γ of 39.76 keV. Instead, AE originate directly from the Rh sample.

The photo-electric attenuation of AE with two different filters reveals the consistent trend using the known data [15], as shown by curve 1 without filters (in red), curve 2 with filters (in blue), and calculated curve 3 (in green) in figures 2(a) and 2(b). However, without including the γ escape peaks, the AE ratios of the calculated value over the experimental data in the band between 24 keV and 28 keV are 0.98±0.02 for Cu filter and 0.91±0.03 for Ta filter, respectively. Here the effective thicknesses of the Cu and Ta filters are respectively 42.6±0.1 μm and 31.68±0.03 μm calibrated by a $^{109}$Cd source with a disk geometry of 2-cm diameter that is comparable to our Rh sample. In contrast, the ratios of the Kα and Kβ peaks (the calculated value over the experimental data) are 0.983±0.001, 0.991±0.002 for the Cu filter and 0.934±0.002, 0.970±0.003 for the Ta filter, which are larger than the AE ratios. It appears that AE are more penetrative than expected. As briefly reported in [1], filters suppress the splitting peaks of K X-rays and γ. Accordingly, an additional photo-electric attenuation of the temporally entangled γ biphoton mimics the anomalous penetration of AE, where the total γ count, including single photon and biphoton, are normalized in the above analysis.

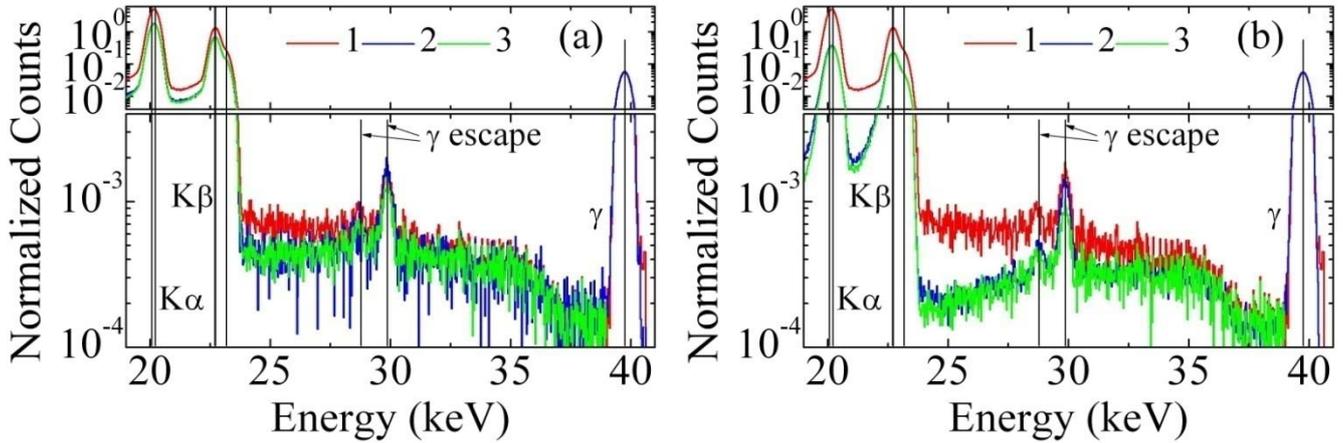

**Fig. 2.** (Colour on line) Normalized spectra, as shown in figure 1 (a), are compared with calculated profiles. Curve 1 in red is the experimental data without filters. (a) Curve 2 in blue is the data with a copper filter. (b) Curve 2 in blue is the data with a tantalum filter. Curve 3 in green are calculated profiles using the photo-electric attenuation of Cu and Ta filters in [15].

Figures 3(a, b, c) show the exponentially-decaying behaviors of γ rays at 39.76 keV (filled circles) and the integrated broad-band count rate from 24 keV to 38 keV (open circles). The corresponding pile-up counts with the decay time constant of 4857/2 s discussed above have been removed by off-line analysis. The solid straight lines in figures 3(a, b, c) are drawn with the theoretical decay time constants of 4857 s. AE are thus identified as directly related to the 39.76 keV γ ray emitted from $^{103m}$Rh nuclei rather than from any other radioactive source excited by bremsstrahlung. The ratio of AE with energy from 24 keV and 38 keV over a single-incoming γ count is 0.292±0.003 in regime III. For further study, the spectra taken in regimes I and II in [3] are also analyzed. The ratios are 0.287±0.005 and 0.292±0.003 in the linear (210 Hz) and nonlinear (295 Hz) regimes in [3], respectively.

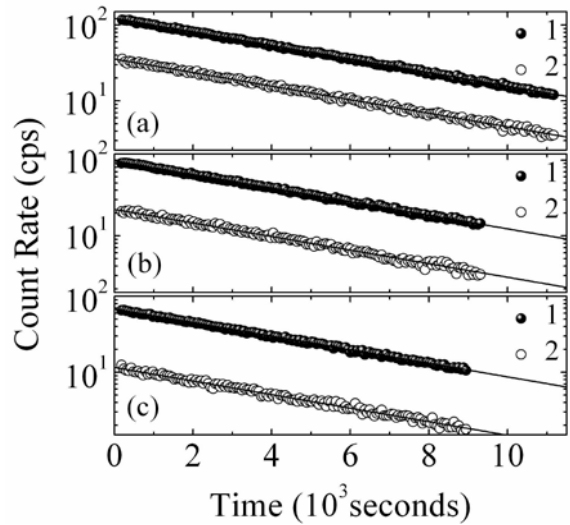

**Fig.3.** Time evolution of the γ count rate at 39.76 keV and the integrated broad-band count rate from 24 keV to 38 keV with and without inserted filters: (a) without filters, (b) with a copper filter, and (c) with a tantalum filter. The solid lines represent the exponential decay of 4857 s. Lines 1 and 2 with filled and open circles correspond to γ and AE, respectively. The counts of pile-ups are removed in off-line data analysis according to parameters obtained from Eq. (1) for (a) and (b). The count rates for (c) are row data without off-line analysis.

Energy spectra below the Kα peak of $^{103m}$Rh and $^{109m}$Ag ($^{109}$Cd source) are shown in figures 4(a) and 4(b) respectively. The intensity of row data is normalized with respect to the total incoming Kα count in both spectra. Since the energy of the

Rh Kα peak (20 keV) is lower than that of Ag Kα peak (22 keV) by about 2 keV, the integrated count from 14 to 16 keV in figure 4(a) corresponds to that from 16 to 18 keV in figure 4(b), by assuming that it is attributed to the Kα peak. By normalization of the spectra over the total count of the Kα peak, any contribution from the process occurring in the detector is not affected by the presence of a filter. This is clearly seen in figure 4(b) with the radioactive source of $^{109m}$Ag ($^{109}$Cd source) with a Cu filter inserted. It is noted that the contribution from Compton scattering corresponding to the γ of 88 keV is negligible in this regime due to the large difference, about two orders of magnitude, in the count rate between these two. With the attenuation by the Cu filter, the emission intensity from 14 to 16 keV in figure 4(a) for $^{103m}$Rh is significantly reduced. This indicates that the AE, discussed extensively in preceding paragraphs, actually extended further down to the energy range below 20 keV. The AE intensity below 20 keV is nearly of order $10^{-4}$ when normalized with respect to the Kα count. It would have been $10^{-2}$, which is larger than the level of the plateau-like continuum above 24 keV by one order of magnitude, if it were normalized with respect to the γ count, since the Kα count is about 86 times larger than the γ count. The difference is attributed to the K-edge absorption of Rh at 23 keV.

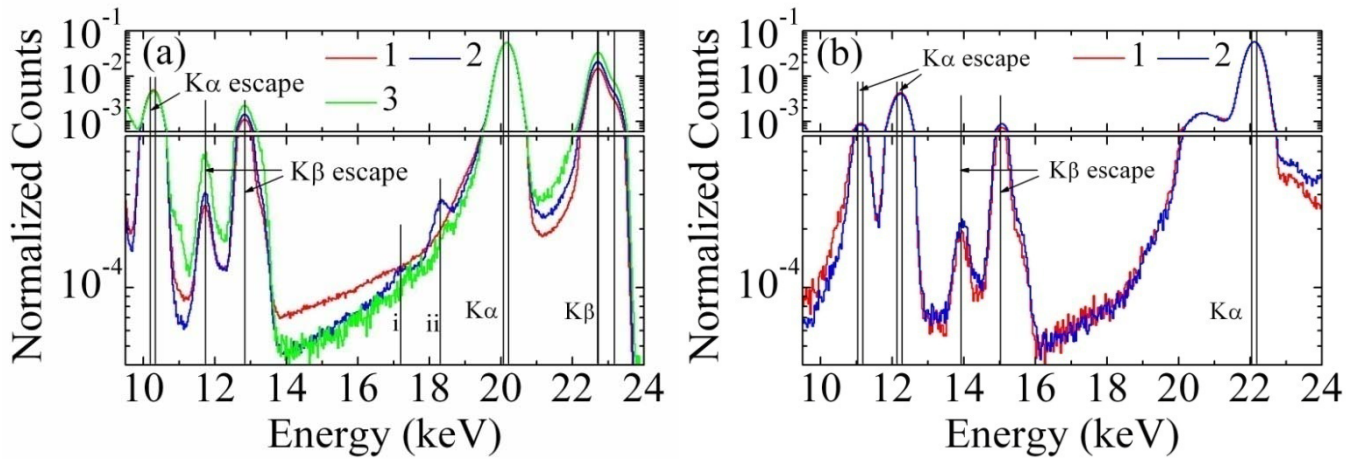

**Fig.4.** (Colour online) Spectra of $^{103m}$Rh and $^{109m}$Ag ($^{109}$Cd source) below Kα peak with and without filters. The spectral profiles are normalized to the incoming Kα count. (a) Spectra of $^{103m}$Rh. Curve 1 in red is without filters. Curve 2 in blue with a Cu filter and curve 3 in green with a Ta filter are lower than curve 1 in the region from 14 to 16 keV. All of the characteristic peaks are identified, including the Kα and Kβ peaks of $^{103m}$Rh with corresponding escapes. Two peaks at (i) 17.2 keV and (ii) 18.3 keV are the coincidence detection elaborated in the text. (b) Spectra of $^{109m}$Ag ($^{109}$Cd). Curve 1 in red is without a Cu filter. Curve 2 in blue is the spectrum with a Cu filter. All characteristic peaks are identified and shown by the vertical lines.

Two peaks surprisingly emerge when a Cu filter was inserted, as shown by the blue line in figure 4(a). One is located at 18.3 keV, marked by the vertical line (ii), and the other at 17.2 keV, marked by the vertical line (i). In fact, these peaks have "single decay speed" verified by the off-line analysis using (1). It is interesting to note that these two peaks are shifted 100 eV to the high energy side with a Ta filter, as indicated by the green line in figure 4(a). This observation reveals that peaks (i) and (ii) with a Cu filter are coincidence counts of Cu Kα peaks at 8.05 keV with Rh Kα escapes at 9.1 and 10.2 keV, respectively, while peaks at 17.3 and 18.4 keV with a Ta filter are the coincidence counts of Ta Lα$_1$ peak at 8.15 keV with Rh Kα escapes at 9.1 and 10.2 keV.

It is worthwhile noting several facts here. First, the same coincidence peaks expected at 19.3 and 20.3 keV are missing for the $^{109}$Cd source with a Cu filter as seen in figure 4(b). Second, coincidence counts of two Cu (Ta) X-rays at 16.1 (16.3) keV are absent in figure 4(a). Third, the 20-keV energy between L and K shells of the Rh atom is unable to create two Ge K holes or even one Ge K hole plus one Ta L hole. Accordingly, one of Ge K X-rays escaping from the detector ejects Cu K and Ta L$_3$ electron respectively in the filters, the emission of which is reabsorbed by the detector. This ping-pong event is enhanced by the creation of double Ge holes, i.e. one K hole and one L hole; otherwise coincidence counts would have shown up with the $^{109}$Cd source. Carefully mapping from figure 4(a) the quotients between curves with and without filters, a peak at 19 keV emerges which undoubtedly supports this long deduction with a single escape of Ge L X-rays. This escape peak is enhanced by the biphoton absorption at the detector surface, e.g., an incident biphoton combination of 8 + 12 keV though the filter windows due to the Cu K edge at 9 keV and the Ta L edge at 10 keV respectively. Hence, we trace the secondary cascade Rh K X-rays back to the primary entangled γ source, discussed briefly in [1, 2]. The K biphoton should have an additional attenuation passing filters, as previously introduced for the γ biphoton. However, this additional attenuation is much weaker than expected for their energies near 10 keV. It seems that the spin-2 biphoton appears weakly-attenuated not only inside the crystal but also outside at the filters.

Conventional wisdom states that the escape is an intrinsic property of the appropriate detector which should not be changed by an external filter unless the K biphoton is additionally attenuated by filters. We carefully check the escape ratio of incoming Kα X-rays. Defined as the ratio of Ge-Kα escape counts with filters to that without filter, the K escape ratios are 0.95 and 0.9 with inserting the Cu and the Ta filters respectively. To further uncover the biphoton entanglement addressed here, a preliminary study with two detectors has shown pronounced coincidence counts. Interestingly, the coincidence counts are likely different in three excitation regimes [1]. Detailed study to resolve their energy distribution is thus required.

The experimental data between 14 and 19.5 keV consist of two energy-dependent contributions, which are the AE distribution $AE(E)$ and the intrinsic detector profile of Kα peak $S_{K\alpha}(E)$. The AE distribution of interest is then resolved by two experimental data, i.e. $S_0(E)$ without filters and $S_{Ta}(E)$ with a Ta filter, channel by channel from the equations

$$S_0(E) = S_{K\alpha}(E) + AE(E)$$
$$S_{Ta}(E) = e^{[\mu(\gamma)-\mu(K\alpha)]d} S_{K\alpha}(E) + e^{[\mu(\gamma)-\mu(E)]d} AE(E)$$   (3)

We apply the photo-electric attenuations $\mu(E)$, $\mu(K\alpha)$ and $\mu(\gamma)$ in [15] to resolve the AE distribution presented in figure 5. The effective thickness $d$ of the filter is obtained by the $^{109}$Cd source as described in the last section. The AE spectral profile shows the left half of the broad-band structure centered around 19.9 keV. Two dips appearing at 17.4 and 18.4 keV are the coincidence counts, as discussed in the preceding paragraph.

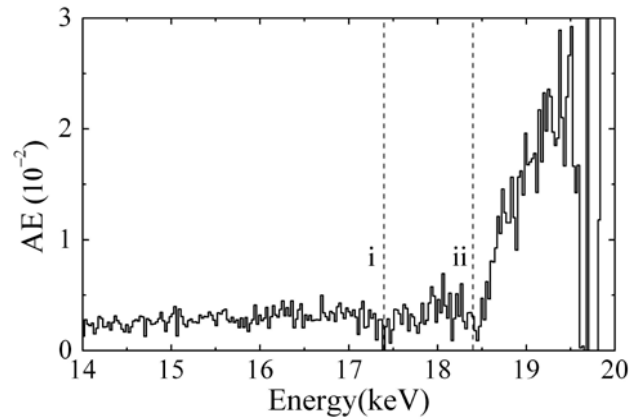

**Fig.5.** AE distribution in the band between 14 and 19.5 keV over the one-incoming γ count calculated by Eq. (3). Two dips are observed at (i) 17.4, and (ii) 18.4 keV, which are attributed to the coincidence counts of the Ta Lα peak at 8.15 keV with Rh Kα escapes at 9.1 and 10.2 keV, respectively. The peak pile-ups between escapes have been removed in off-line analysis using Eq. (1).

## 4. Conclusion

The AE distribution centered on the one half transition energy is one of the key indications for the biphoton transition. Another is the coincidence counts appearing at the same detector. Due to the presence of Rh K lines near 20 keV and their escapes near 10 keV, the experimental search on central AE peaks of the nuclear cascade at 20 keV and the atomic cascade at 10 keV is much more difficult for $^{103m}$Rh than for $^{93m}$Nb and $^{45m}$Sc. The AE distribution below 20 keV increases toward 20 keV, which shows a bump above 18 keV in figure 5. The AE distribution above 20 keV also increases toward 20 keV, while the bump above 20 keV is buried under Rh K lines. We have worked on these data for years. Several open questions remain unsolved, *e.g.,* whether AE is symmetrically distributed about the half transition energy and whether the biphoton radiation is coherent outside the crystal.

**Acknowledgments**

Y. Cheng gives special thanks to Chinping Chen for proof-reading the experimental part of the text and G. E. Volovik for a discussion on the magnon BEC under parametric pumping. This work was supported by the NSFC grant 10675068.